\documentclass[12pt]{article}

\usepackage{epsfig}

\usepackage{amsmath}
\usepackage{amssymb}
\usepackage{euscript}

\newcommand{\Ucal}{{\cal U}}
\newcommand{\Peu}{\EuScript{P}}
\newcommand{\Keu}{\EuScript{K}}

\newcommand{\veps}{\varepsilon}

\newcommand{\tG}{{\bf G}'}

\begin{document}

\begin{titlepage}


\begin{flushright}
\bf IFJPAN-V-04-09
\end{flushright}

\vspace{1mm}
\begin{center}
  {\Large\bf%
    Hierarchically Organized Iterative
    Solutions of the Evolution Equations in QCD$^{\star}$
}
\end{center}
\vspace{3mm}

\begin{center}
{\bf S. Jadach, M. Skrzypek}
{\em and}
{\bf Z. W\c{a}s} \\

\vspace{1mm}
{\em Institute of Nuclear Physics, Polish Academy of Sciences,\\
  ul. Radzikowskiego 152, 31-342 Cracow, Poland,}\\
\end{center}

\vspace{8mm}
\begin{abstract}
The task of Monte Carlo simulation of the evolution of the parton distributions
in QCD and of constructing new parton shower Monte Carlo algorithms
requires new way of organizing solutions of the QCD evolution equations,
in which quark$-$gluon transitions on one hand and
quark$-$quark or gluon$-$gluon transitions (pure gluonstrahlung)
on the other hand, are treated separately and differently.
This requires certain reorganization of the iterative
solutions of the QCD evolution equations and leads to
what we refer to as a {\em hierarchic iterative solutions}
of the evolution equations.
We present three formal derivations of such a solution.
Results presented here are already used in the other recent works
to formulate new MC algorithms for
the parton-shower-like implementations of the QCD evolution equations.
They are primarily of the non-Markovian type.
However, such a solution can be used for the Markovian-type MCs as well.
We also comment briefly on the relation of the presented formalism
to similar methods used in other branches of physics.
\end{abstract}

\vspace{4mm}
\begin{center}
\em To be submitted to Acta Physica Polonica
\end{center}

\vspace{15mm}
\begin{flushleft}
{\bf IFJPAN-V-04-09\\
     December~2006}
\end{flushleft}

\vspace{5mm}
\footnoterule
\noindent
{\footnotesize
$^{\star}$This work is partly supported by the EU grant MTKD-CT-2004-510126
 in partnership with the CERN Physics Department and by the Polish Ministry
 of Scientific Research and Information Technology grant No 620/E-77/6.PR
 UE/DIE 188/2005-2008.
}

\end{titlepage}

\section{Introduction to the problem}

The standard QCD evolution equation
\begin{equation}
\label{eq:Evolequ}
\begin{split}
  \partial_t D_k(t,x)
&= \sum_j {\Peu}_{kj}(t,\cdot)\otimes D_j(t,\cdot)(x),
\\
f(\cdot){\otimes} g(\cdot)(x) 
   &\equiv \int_0^1 dx_1 dx_2 \delta(x-x_1 x_2)f(x_1)g(x_2)
    \equiv \int_x^1  \frac{dx_2}{x_2} f
                      \Bigl(\frac{x}{x_2}\Bigr)g(x_2),
\end{split}
\end{equation}
describes response of the parton distribution function (PDF) $D_k(t,x)$
to a change of the large energy scale $Q=\exp(t)$.
Variable $x$ is identified as a fraction of the hadron momentum carried by parton
of the type $k$ (quark, gluon). 
Evolution kernel ${\Peu}$ is calculable within perturbative QCD.

The above evolution equation (\ref{eq:Evolequ})
is an important ingredient in many QCD perturbative calculations.
It can be solved using variety of the numerical methods, including Monte Carlo method.
The knowledge of $D_{k}(t_0,x)$ at certain initial $t_0$, is required 
for solving evolution equation at other $t>t_0$.
The initial PDF is fitted to experimental data.

\begin{figure}[!ht]
  \centering
  \epsfig{file=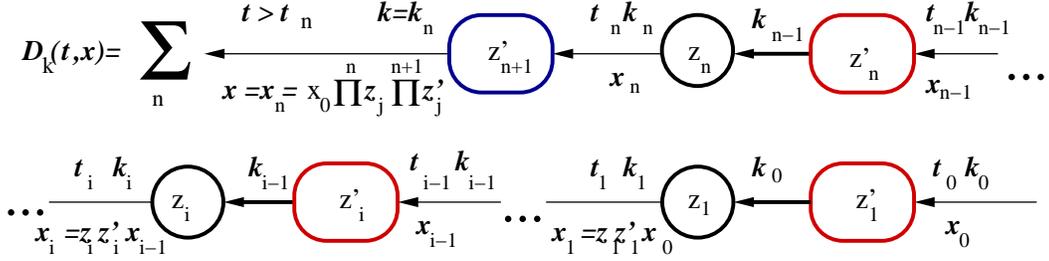,width=140mm}
  \caption{\sf
	Kinematics in the 2-level
	emission tree of eq.~(\ref{eq:Evolsolu}).
    }
  \label{fig:hierarchy}
\end{figure}

The principal aim of this work is to derive the following%
 analytical solution of the
above QCD evolution equations
\begin{equation}
\label{eq:Evolsolu}
\begin{split}
D_k(t,x) =&
     \int\limits_0^1 dz'\; dx_0\; 
     G_{kk}^B(t, t_{0},z')\;
     D_k(t_0,x_0)\delta(x-z' x_0)
  +\sum_{n=1}^\infty \;
      \int\limits_0^1 dx_0\;
\\&~~~~~~\times
   \sum_{{k_{n-1},\dots,k_{1},k_{0}}}
   \Bigg[ \prod_{j=1}^n \int\limits_{t_0}^t dt_j\;
     \Theta(t_j-t_{j-1}) \Bigg]\;
   \int\limits_0^1 dz'_{n+1}\;
   \Bigg[ \prod_{i=1}^{n} 
     \int\limits_0^1 dz'_i\;dz_i\;
     \Bigg]\;
\\&~~~~~~\times
   G_{kk}^B(t,t_n,z'_{n+1})\;
   \Bigg[ \prod_{i=1}^n
     \Peu_{k_ik_{i-1}}^A (t_i,z_i)\;
     G_{k_{i-1}k_{i-1}}^B(t_i,t_{i-1},z'_i) \Bigg]\;
\\&~~~~~~\times
      D_{k_0}(t_0,x_0) 
      \delta\bigg(x- x_0 \prod_{i=1}^n z_i \prod_{i=1}^{n+1} z'_i
     \bigg),
   \;\;\; k_n=k,
\end{split}
\end{equation}
where we have isolated flavour conserving (bremsstrahlung) part of the kernel%
\footnote{Since $\Peu^A_{kk}=0$, only flavour changing indices
$k_i\neq k_{i-1}$ enter in the flavour sum in eq.~(\ref{eq:Evolsolu}).}
$\Peu^B_{kj}\equiv\delta_{kj}\Peu_{kk}$ from
the total kernel, $\Peu_{kj}=\Peu^A_{kj}+\Peu^B_{kj}$,
and $G_{kk}^B(t_1,t_0,z)$ is the solution of the following simplified evolution
equation, similar to eq.~(\ref{eq:Evolequ}),
\begin{equation}
\label{eq:EvolGA}
 \partial_t G_{kk}^B(t,t_0,z) = 
    {\Peu}^B_{kk}(t,\cdot)\otimes G_{kk}^B(t,t_0,\cdot)(z),
\end{equation}
see Section \ref{sec:real-thing} for the details.
The boundary condition is $G_{kk}^B(t_0,t_0,z)=\delta(1-z)$.
See also fig.~\ref{fig:hierarchy} for graphical representation.

It is important to provide formal proof of eq.~(\ref{eq:Evolsolu}),
because it is a critical ingredient in several new Monte Carlo algorithms
of the non-Markovian type%
\footnote{It can serve as basis of a novel type of the Markovian MC as well.}
described in refs.~\cite{Golec-Biernat:2006xw} and~\cite{Jadach:2003bu},
and possibly in other future works.

Let us now explain in details
notation used in eqs.~(\ref{eq:Evolequ}-\ref{eq:EvolGA}).
In function $D_k(t,x)$
variable $1\geq x \geq 0$ is the fraction 
of the hadron momentum carried by the parton
of the type $k=G,q_i,\bar{q}_i$, i.e. gluon, quark or antiquark,
at the high energy scale $Q$, conveniently translated into 
the ``evolution time'' variable $t=\ln Q$.
In QCD the PDF represents the wave function
of the hadron close to the light-cone.
See ref.~\cite{Collins:2003fm} for an expert discussion on the precise 
meaning of PDF in QCD, in a wide context of the so-called
{\em factorization theorems}~\cite{Collins:1984kg,Collins:1981uk,Bodwin:1984hc}
in the gauge Quantum Field Theories.

\begin{figure}[!ht]
  \centering
  {\epsfig{file=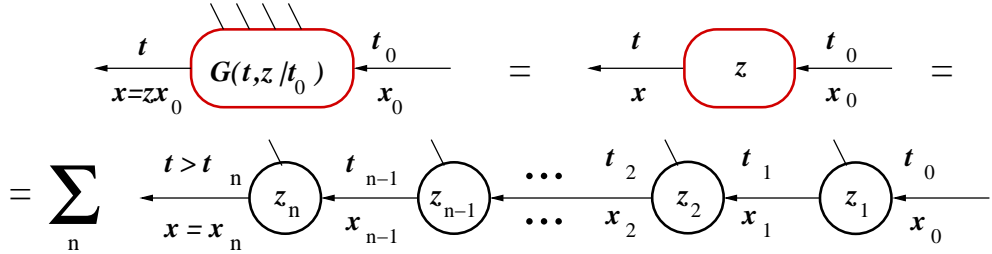,width=140mm}}
  \caption{\sf
	Kinematics in the
	pure brems\-strahlung emission tree of eq.~(\ref{eq:GAbrems}).
    }
  \label{fig:bremstree}
\end{figure}

In this work we shall restrict ourselves to the most
common QCD evolution equations of the DGLAP type~\cite{DGLAP},
with the kernel splitting functions%
\footnote{The DGLAP kernels in the $\overline{MS}$ scheme were
   calculated in QCD at the two levels beyond the 
   leading-logarithmic (LL) approximation.}
incorporating the QCD coupling constant 
(for the sake of the simplicity of notation)
\begin{equation}
  {\Peu}_{kj}(t,z) = \frac{\alpha(t)}{\pi} P_{kj}(t,z).
\end{equation}
The QCD kernel functions are singular,
with singularities of the type $(\ln(1-z)^n/(1-z))_+$.
We shall typically regularize them with the help of an explicit 
small infrared (IR) cutoff parameter%
\footnote{The infinitesimal parameter $\veps$ can be $t$-dependent,
  without any loss of generality in the following treatment.}
$\veps$ as follows:
\begin{equation}
\label{Peu}
{\Peu_{kj}(t,z)}=-\Peu^{\delta}_{kk}(t,\veps) \delta_{kj}\delta(1-z)
                 +\Peu^{\Theta}_{kj}(t,z),
\;\;
       \Peu^{\Theta}_{kj}(t,z)=\Peu_{kj}(t,z)\Theta(1-z-\veps).
\end{equation}
The important Sudakov formfactor $\Phi_{k}(t,t_0)$
is directly related to the virtual part of the kernels:
\begin{equation}
    {\Phi_{k}(t,t_0)} = \int\limits_{t_0}^t  dt'\;
       \Peu^\delta_{kk}(t',\veps).
\end{equation}
Finally the (bremsstrahlung-type) auxiliary distribution 
\begin{equation}
\begin{split}
\label{eq:GAbrems}
G_{kk}^B &(t,t_0,z) = \delta(1-z)
\\ &
  +\sum_{n=1}^\infty \;
      \bigg[ \prod_{i=1}^n \int\limits_{t_0}^t dt_i\;
      \Theta(t_i-t_{i-1})  \int\limits_0^1 dz_i\bigg]
      e^{-\Phi_k(t,t_n)}
      \bigg[\prod_{i=1}^n 
           \Peu_{kk}^\Theta (t_i,z_i) 
            e^{-\Phi_{k}(t_i,t_{i-1})} \bigg]
      \delta\big(x- \prod_{i=1}^n z_i \big)
\end{split}
\end{equation}
is an iterative%
\footnote{We shall explain in the next Section why we call it ``iterative''.}
solution of the flavour-diagonal evolution equation of eq.~(\ref{eq:EvolGA}).
See also fig.~\ref{fig:bremstree} for graphical representation of the
above gluonstrahlung segment of the evolution.

\section{The solutions}
If our only aim was to prove the correctness of eq.~(\ref{eq:Evolsolu})
as a solution of eq.~(\ref{eq:Evolequ}),
then the simplest approach would be just to substitute it into this equation
and check with a little bit of algebra that indeed it is the solution.
Our aims are however more general:
(i) to derive eq.~(\ref{eq:Evolsolu}) in a more systematic way,
(ii) to understand better its relation to the other widely known
and used iterative solutions of eq.~(\ref{eq:Evolequ}),
(iii) to prove that its {\em exclusive} content,
in terms of the fully differential distribution 
in all variables $t_i$ and $z_i$, $i=1,2,...,n$, for each $n$,
is exactly the same as in other iterative solutions, commonly used in the
MC approaches.

Having all the above in mind, let us proceed methodically,
first with deriving solution of the evolution of eq.~(\ref{eq:Evolequ}),
in terms of a time-ordered exponential, widely used in the literature.
Next, we shall present first example of the derivation of eq.~(\ref{eq:Evolsolu})
by means of reorganizing the evolution equation and solving it once again.
Then, we shall present second example of the derivation,
in which the above time-ordered exponential
is algebraically reorganized (transformed) into eq.~(\ref{eq:Evolsolu}).
Finally the third derivation of eq.~(\ref{eq:Evolsolu}) based on straightforward
reorganization of the multiple sums and integrals will be included in the Appendix.

\subsection{Time-ordered exponential}

The solution of eq.~(\ref{eq:Evolequ})
can be established quickly and rigorously,
for instance by means iteration, as a time-ordered exponential
of the kernel operator $\Peu$ in the vector (linear) space indexed
by one continuous variable $x$ and one discrete $k$.
More precisely,
eq.~(\ref{eq:Evolequ}) in a more compact matrix notation reads
\begin{equation}
  \partial_t {\bf D}(t) = {\bf P}(t) \; {\bf D}(t)
\end{equation}
and its solution in the same compact matrix notation is given by
\begin{equation}
\label{eq:solumatr} 
  {\bf D}(t) = \exp\left( \int_{t_0}^t {\bf P}(t') dt' \right)_T \; {\bf D}(t_0)
       = {\bf G}_{P}(t,t_0) D(t_0),
\end{equation}
where we employ the following well known {\em time-ordered exponential}
evolution operator%
\footnote{Here and in the following we define 
$\prod_{i=1}^n A_i \equiv A_n A_{n-1}\dots A_2A_1$.}
\begin{equation}
\label{eq:torder}
{\bf G}_{H}(t,t_0)=
{\bf G}(H;t,t_0)=
\exp\left( \int_{t_0}^t {\bf H}(t') dt' \right)_T
 = {\bf I} +\sum_{n=1}^\infty 
      \prod_{i=1}^n  \int_{t_0}^t dt_i \theta_{t_i>t_{i-1}} {\bf H}(t_i),
\end{equation}
for $t\geq t_0$.
It is familiar to all readers
from the textbooks of the Quantum Mechanics%
\footnote{However, we do not require ${\bf H}$ to
     be hermitian and ${\bf G}$ to be unitary.}.
We define function $\theta_{x>y}$ 
to be equal 1 when $x>y$ and equal 0 otherwise.

The compact solution of equation (\ref{eq:solumatr}),
can be translated into more traditional integro-tensorial notation,
with explicit sums and integrals:
\begin{equation}
  \label{eq:IterPeu}
  \begin{split}
 D_k(t,x)
 &= \left\{
   \left[ \exp\left( \int_{t_0}^t dt'\; {\bf P}(t',\cdot)\otimes 
   \right)\right]_T\; {\bf D}(t_0,\cdot)
   \right\}_k
\\&
= D_k(t_0,x)
  +\sum_{n=1}^\infty \;
   \sum_{k_{n-1}...k_{1}k_{0}}
   \int_0^1 dx_0\;
\\&~~~~~~~~~~~~~~~~~\times
   \bigg[ \prod_{j=1}^n \int\limits_{t_0}^t dt_j\; 
          \theta_{t_j>t_{j-1}} 
          \int\limits_0^1 dz_j\;
          \Peu_{k_jk_{j-1}}(z_i)
   \bigg]\;
     D_{k_0}(t_0,x_0)\; \delta_{x= x_0\prod_{i=1}^n z_i},
  \end{split}
\end{equation}
where $\delta_{x=y}\equiv \delta(x-y)$.
We realize, that the above formula, although rigorous and very elegant,
is useless for any practical evaluations,
because it contains polynomials of the negative and singular terms
comming from products of $-\Peu^\delta \delta(1-z)$ factors.
In principle these inconvenient terms can be resummed
(exponentiated) with the help of the direct but tedious algebra
on the multiple flavour indices and $z$-integrals%
\footnote{Similarly as the method used in the Appendix to re-sum
another part of the kernel.}.
In the following section we shall do it using more elegant methods.

Having defined the time ordered exponential evolution operator ${\bf G}_{H}$,
let us quote its basic features and extend its definition for the latter use.
The well known rule%
\footnote{Chapman-Kolmogorov-Smoluchowski-Einstein relation,
see ref.~\cite{kampen,walecka}.}
\begin{equation}
\label{eq:Urule}
   {\bf G}_{H}(t,t_x){\bf G}_{H}(t_x,t)= {\bf G}_{H}(t,t_0),
   \quad t\geq t_x \geq t_0
\end{equation}
helps to manipulate products  of the time-ordered exponents.
We may also define the inverse operator for the 
``backward evolution'' ($t<t_0$) as follows
\begin{equation}
\label{eq:BackEvol}
  {\bf G}_H(t_0,t) \equiv {\bf G}^{-1}_{H}(t,t_0),\quad t<t_0,
\end{equation}
where the inverse operator is constructed using%
\footnote{
  The algebraic proof that 
  ${\bf G}^{-1}_{H}(t,t_0){\bf G}_{H}(t,t_0)={\bf I}$,
  using eq.~(\ref{eq:torder}), we leave to the reader.
  The matrix elements of ${\bf G}^{-1}$
  can be non-positive and highly singular.}
${\bf G}^{-1}_{H}(t,t_0)\equiv {\bf G}_{(-H)}(t,t_0)$.
With help of the above definition
validity of eq.~(\ref{eq:Urule}) can be extended to any $t_x$.

\subsection{Derivation by reorganizing evolution equation}

In the following we show how to resume singular $\Peu^\delta$ terms
by going back to the evolution equation,
reorganizing  it and solving it once again.
We are going to show this standard trick in a detail,
because, subsequently, we shall generalize it to the case
of an arbitrary part of the kernel (instead of the $\Peu^\delta$ part).
It is essentially a warm-up example.

\subsection{Resuming virtual part of the kernel -- warm-up example}
Inserting explicitly regularized kernel, our evolution equation 
takes the following form
\begin{equation}
  \label{eq:EvolD}
  \partial_t D_k(t,x)=
  - \Peu^{\delta}_{kk}(t)\; D_k(t,x)
  +\sum_j \Peu^{\Theta}_{kj}(t,\cdot)\otimes D_j(t,\cdot)(x).
\end{equation}
It can be transformed into%
\footnote{This integro-differential form is exposed in the
 QCD textbooks, see ref.~\cite{stirling-book},
 and is also used in the numerical
 evaluation (evolution) of PDFs using non-Monte-Carlo
 methods, for instance ref~\cite{qcdnum16}.}
\begin{equation}
  \label{eq:EvolTilde}
    \partial_t \Big(  e^{\Phi_k(t,t_0)} D_k(t,x) \Big)
    = \sum_j 
      e^{\Phi_k(t,t_0)}  \Peu^\Theta_{kj}(t,\cdot) e^{-\Phi_j(t,t_0)}
      \otimes e^{\Phi_j(t,t_0)}  D_j(t,\cdot)(x).
\end{equation}
Changing slightly notation the above is transformed into
\begin{equation}
\begin{split}
  &\partial_t \tilde{D}_k(t,x)
    = \sum_j \tilde\Peu^\Theta_{kj}(t,\cdot) \otimes \tilde{D}_j(t,\cdot)(x),
\\&
   \tilde{D}_k(t,x) \equiv \exp(\Phi_k(t,t_0)) D_k(t,x),
\\&
  \tilde\Peu^\Theta_{kj}(t,z) \equiv
  \exp(\Phi_k(t,t_0))  \Peu^\Theta_{kj}(t,z) \exp(-\Phi_j(t,t_0))
\end{split}
\end{equation}
or in an equivalent compact matrix formulation it reads
\begin{equation}
\begin{split}
   \partial_t \tilde{D}(t)= \tilde{\bf P}^{\Theta}(t) \; \tilde{D}(t).
\end{split}
\end{equation}
The time ordered solution
\begin{equation}
\label{eq:soluP}
\begin{split}
\tilde{D}(t)&= \exp\left( \int_{t_0}^t \tilde{\bf P}^{\Theta}(t') dt'\right)_T\;
               \tilde{D}(t_0)
             = {\bf G}_{\tilde{\bf P}^\Theta}(t,t_0)\; \tilde{D}(t_0)
\end{split}
\end{equation}
of the evolution equation is widely known and exploited routinely in
many practical evaluations of solutions of the QCD evolution.
It is usually written in the traditional integro-tensorial
representation similarly as eq.~(\ref{eq:IterPeu}),
in terms of of the initial $D(t_0)$ and the product of $\Peu^\Theta$,
taking the following familiar shape:
\begin{equation}
\label{eq:SolIter}
\begin{split}
 &D_k(t,x) = e^{-\Phi_k(t,t_0)} D_k(t_0,x)
  +\sum_{n=1}^\infty \;
   \sum_{k_0,\dots,k_{n-1}}
      \bigg[ \prod_{i=1}^n \int\limits_{t_0}^t dt_i\;
      \Theta(t_i-t_{i-1})  \int\limits_0^1 dz_i\bigg]
\\&~~~~~~~\times
      e^{-\Phi_k(t,t_n)}
      \int\limits_0^1 dx_0\;
      \bigg[\prod_{i=1}^n 
           \Peu_{k_ik_{i-1}}^\Theta (t_i,z_i) 
            e^{-\Phi_{k_{i-1}}(t_i,t_{i-1})} \bigg]
     D_{k_0}(t_0,x_0) \delta\big(x- x_0\prod_{i=1}^n z_i \big).
  \end{split}
\end{equation}
Trivial identity
\begin{equation}
\label{eq:trivial}
e^{-\Phi_i(t_i,t_0)} e^{\Phi_i(t_{i-1},t_0)}= e^{-\Phi_i(t_i,t_{i-1})}
\end{equation}
was also employed.
The above solution of the evolution equation
is used as a basic formula in the Monte Carlo evaluation of the PDFs
using Markovian MC algorithms,
see for example ref.~\cite{Jadach:2003bu}.
We shall refer to this solution as non-hierarchic iterative solution of the
evolution equation.

Let us remind the reader, that
in the above warm-up exercise we have resummed the relatively simple component 
$\Peu^B_{jk}(t,z)=-\delta_{jk}\Peu^\delta_{kk}(t)\delta(1-z)$
of the evolution kernel,
which was completely diagonal, both in $k$ and in $z$.
In this special case ${\bf G}^{-1}_{\Peu^B}$
is trivially calculable, contrary to more general case of non-diagonal
$\Peu^B$ discussed in the following.

Let us now come back to our principal aim,
proving eq.~(\ref{eq:Evolsolu}), where less trivial component of the
kernel will be isolated/resummed.

\subsection{Resumming gluonstrahlung -- the real thing}
\label{sec:real-thing}

In order to prove eq.~(\ref{eq:Evolsolu}),
we need to resum (exponentiate) the following part of the kernel
\begin{equation}
  \Peu_{jk}^B(t,z) = \delta_{jk} \Peu_{kk}(t,z)
                   =  -\delta_{jk}\Peu^\delta_{kk}(t)\delta(1-z)
                      +\delta_{jk}\Peu^\Theta_{kk}(t,z),
\end{equation}
which is diagonal in the flavour indices, but not in $z$.
This part of the kernel is always IR divergent and generates
multiple gluon emission process, that is {\em gluonstrahlung}.
The remaining {\em flavour-changing}
part of the full kernel is defined as $\Peu^A=\Peu-\Peu^B$.
The original full evolution equation and its solution read
\begin{equation}
 \partial {\bf D}(t)= \Big({\bf P}^A(t) + {\bf P}^B(t)\Big)\;  {\bf D}(t),\quad
 D(t) = \exp\left( \int_{t_0}^t \Big({\bf P}^A(t')+{\bf P}^B(t')\Big) dt' 
           \right)_T \; {\bf D}(t_0).
\end{equation}
At this point, the $G_A$-function of eq.~(\ref{eq:GAbrems})
can be identified with the following operator
\begin{equation}
\label{eq:GA}
 {\bf G}_B(t,t_0) \equiv {\bf G}_{P^B}(t,t_0)
  = \exp\left(\int_{t_0}^t {\bf P}^B(t')dt' \right)_T,\quad
\end{equation}
where
\begin{equation}
\label{eq:GA2}
 \partial_t {\bf G}_B(t,t_0) = {\bf P}^B(t)\; {\bf G}_B(t,t_0),
\end{equation}
see also eq.~(\ref{eq:EvolGA}).

In order to derive eq.~(\ref{eq:Evolsolu}) we proceed
analogously as in the derivation of eq.~(\ref{eq:SolIter});
we shall introduce ${\bf G}_A(t,t_0)$ in the evolution equation,
similarly as we have introduced $\exp\big(\Phi_k(t,t_0)\big)$.
The main complication will be in the non-commutative
nature of ${\bf G}_A(t,t_0)$.
Let us introduce in the evolution an equation auxiliary PDF
\begin{equation}
  \tilde{\bf D}(t) = {\bf G}^{-1}_B(t,t_0)\; {\bf D}(t),\quad
   {\bf D}(t)={\bf G}_B(t,t_0)\;\tilde{\bf D}(t),
\end{equation}
getting after the differentiation
\begin{equation}
  {\bf G}_A(t,t_0)\; \partial \tilde{D}(t)
 +\Big(\partial{\bf G}_A(t,t_0)\Big) \tilde{D}(t)
= \Big({\bf P}^A(t) + {\bf P}^B(t)\Big)\; {\bf G}_A(t,t_0)\;\tilde{D}(t).
\end{equation}
After inserting eq.~(\ref{eq:GA2}) we obtain
\begin{equation}
  {\bf G}_B(t,t_0)\; \partial \tilde{\bf D}(t)
 +{\bf P}^B(t)\; {\bf G}_B(t,t_0) \tilde{\bf D}(t)
= \Big({\bf P}^A(t) + {\bf P}^B(t)\Big)\; {\bf G}_A(t,t_0)\;\tilde{\bf D}(t).
\end{equation}
The term proportional to ${\bf P}^A$ gets eliminated
\begin{equation}
  {\bf G}_B(t,t_0)\; \partial \tilde{\bf D}(t)
= {\bf P}^A(t)\; {\bf G}_B(t,t_0)\;\tilde{\bf D}(t).
\end{equation}
and we return to the usual evolution equation
\begin{equation}
\partial \tilde{\bf D}(t)
  = \tilde{\bf P}^A(t)\;\tilde{\bf D}(t),\qquad
\tilde{\bf P}^A(t)
\equiv {\bf G}^{-1}_B(t,t_0)\; {\bf P}^A(t)\;{\bf G}_B(t,t_0)\;
\end{equation}
with the usual solution
\begin{equation}
 {\bf D}(t) = {\bf G}_B(t,t_0)\;
  \exp\left( \int_{t_0}^t \tilde{\bf P}^A(t') dt' \right)_T \;
  {\bf D}(t_0).
\end{equation}

The last step on the way to eq.~(\ref{eq:Evolsolu})
is elimination of the operator ${\bf G}^{-1}_B(t,t_0)$ being
part of $\tilde{\bf P}$.
The reason for that is that ${\bf G}^{-1}$ is not well suited for
any numerical evaluation, especially of the MC type, due to alternating sign
in the exponential expansion, hence it is better to eliminate it
from the final result.
It is done with the help of the following identity
\begin{equation}
\begin{split}
&{\bf G}_B(t,t_0)\; \exp\left( \int_{t_0}^t \tilde{\bf P}^A(t') dt' \right)_T=
\\&~~~~~~~~~~~~~
={\bf G}_B(t,t_0)+
 \sum_{n=1}^\infty
 \left[
 \prod_{i=1}^n \int_{t_0}^{t}dt_i\; \theta_{t_i>t_{i-1}}
    {\bf G}_B(t_{i+1},t_{i})
    {\bf P}^A(t_{i})
 \right]
 {\bf G}_B(t_{1},t_{0}),
\end{split}
\end{equation}
where $t_{n+1}\equiv t$.
This identity is derived rather easily by inspecting each $n$-th term in the
expansion of the time ordered exponent and applying
the following relation%
\footnote{See eqs.~(\ref{eq:Urule}-\ref{eq:BackEvol}) 
       and the accompanying discussion.}
(analogous to eq.~(\ref{eq:trivial}))
\begin{equation}
{\bf G}_B(t_{i+1},t_0)\; {\bf G}^{-1}_B(t_i,t_0)
={\bf G}_B(t_{i+1},t_{i})\; {\bf G}_B(t_i,t_0)\; {\bf G}^{-1}_B(t_i,t_0)
= {\bf G}_B(t_{i+1},t_{i})
\end{equation}
for each pair ${\bf G}_B {\bf G}^{-1}_B$ sandwiched between
adjacent  ${\bf P}^A$'s.
The final solution reads
\begin{equation}
\label{eq:solumatr2}
\begin{split}
&{\bf D}(t)=
{\bf G}_B(t,t_0)\; {\bf D}(t_0)+
\\&~~~~~~~~~~~~~
+\sum_{n=1}^\infty
 \left[
 \prod_{i=1}^n \int_{t_0}^{t}dt_i\; \theta_{t_i>t_{i-1}}
    {\bf G}_B(t_{i+1},t_{i})
    {\bf P}^A(t_{i})
 \right]
 {\bf G}_B(t_{1},t_{0})\; {\bf D}(t_0).
\end{split}
\end{equation}
When translated into the integro-tensorial notation, the above formula
turns out to be identical with our target eq.~(\ref{eq:Evolsolu}).
In this way we have completed its derivation.

It is now obvious why
eq.~(\ref{eq:Evolsolu}) we call a {\em hierarchic} solution of the
evolution equation.
It is because its components ${\bf G}_A$ are solutions of 
another simpler evolution equation (gluonstrahlung) of its own.
Higher level evolution embeds lower level simpler evolution as
a building block.

\subsection{Derivation by reorganizing time-ordered exponential}
A disadvantage of the derivation presented above is that
it exploits the inverse evolution operator ${\bf G}^{-1}$,
which is in a general case difficult to define properly, while it drops out from
the final result of eqs.~(\ref{eq:Evolsolu}) 
or (\ref{eq:solumatr2}) anyway.
The natural question is therefore whether we could derive 
eq.~(\ref{eq:Evolsolu}) without introducing
the operator ${\bf G}^{-1}$ in the intermediate stages of the proof.

Furthermore, going back to the modified evolution equation
and solving it once again obscures the relation%
\footnote{Such a relation is relevant for parton shower applications.}
between variables $(k_i,z_i)$
in the non-hierarchic solution of eq.~(\ref{eq:SolIter}) on one hand
and the hierarchic one of eq.~(\ref{eq:Evolsolu}) on the other hand.
In the following we shall, therefore, 
present an alternative example of the derivation of eq.~(\ref{eq:solumatr})
without explicit use of the inverse evolution operator ${\bf G}^{-1}$.
In such a case, the relation between variables $(k_i,z_i)$
in eq.~(\ref{eq:SolIter}) and eq.~(\ref{eq:Evolsolu})
can be traced back (recovered) more easily.

The following derivation will be strongly reminiscent to a derivation
of identity $F(x,y)=\exp(x+y)=\exp(x)\exp(y)$ 
by means of the Taylor expansion%
\footnote{Note that such a derivation is almost equivalent
to a direct multiplication of the infinite sums for $\exp(x)$ and $\exp(y)$,
but more transparent algebraically.}
with respect $y$ i.e.
$F(x,y)= \sum_{n=0}^\infty (x^n/n!) \partial^n_x F(x,y)\big|_{x=0}$.

Let us introduce slightly modified evolution operator
\begin{equation}
\tG_{\bf H}(t,t_0)=
\tG({\bf H};t,t_0)= {\bf G}({\bf H};t,t_0)\;\theta_{t\geq t_0},
\end{equation}
where ${\bf G}$ was already defined as the time-ordered exponential
in eq.~(\ref{eq:torder}).
The additional $\theta$-factor ensuring $t\geq t_0$ is
will make the following algebra more compact.
We define ${\bf H}={\bf H}_\lambda={\bf B}+\lambda {\bf A}$
(we shall set $\lambda=1$ at the very end of calculation).
The whole derivation relies on the following identity%
\footnote{Quite similar identity holds for an arbitrary 
   $\lambda$-dependence in ${\bf H}(\lambda)$.}
\begin{equation}
\label{eq:Urule3} 
  \partial_\lambda \tG_{\bf H}(t,t_0)
 =  \int_{t_0}^{t} dt_1\;
    \tG_{\bf H}(t,t_1)\; {\bf A}(t_1)\;
    \tG_{\bf H}(t_1,t_0),
\end{equation}
which can be derived using definition 
of eq.~(\ref{eq:torder}), 
and reorganizing all integrations over $t_i$'s.
The second derivative follows trivially:
\begin{equation}
\begin{split}
  \partial^2_\lambda \tG_{\bf H}(t,t_0)
&=  \int_{t_0}^{t} dt_1\; dt_2\;
    \tG_{\bf H}(t,t_2)\;   {\bf A}(t_2)\;
    \tG_{\bf H}(t_2,t_1)\; {\bf A}(t_1)\;
    \tG_{\bf H}(t_1,t_0)\\
&+  \int_{t_0}^{t} dt_1\; dt_2\;
    \tG_{\bf H}(t,t_1)\;   {\bf A}(t_2)\;
    \tG_{\bf H}(t_1,t_2)\; {\bf A}(t_1)\;
    \tG_{\bf H}(t_2,t_0)\\
&=2!\int_{t_0}^{t} dt_2\; dt_1\;
    \tG_{\bf H}(t,t_2)\;   {\bf A}(t_2)\;
    \tG_{\bf H}(t_2,t_1)\; {\bf A}(t_1)\;
    \tG_{\bf H}(t_1,t_0),
\end{split}
\end{equation}
and the $n-$th derivative is%
\footnote{Strictly speaking we should use mathematical induction 
   over $n$ to verify this.}
\begin{equation}
  \partial_\lambda^n \tG_{{\bf H}_\lambda}(t,t_0)
 =  n!\;\prod_{i=1}^n \Bigg(
    \int_{t_0}^{t} dt_i\;
    \tG_{{\bf H}_\lambda}(t_{i+1},t_{i})\; {\bf A}(t_{i})\;
    \Bigg)
    \tG_{{\bf H}_\lambda}(t_1,t_0),
\end{equation}
where $t_n\equiv t$.
Now, let us use Taylor expansion
\begin{equation}
\tG_{{\bf H}_\lambda}(t,t_0)
=\tG_{{\bf H}_{\lambda=0}}(t,t_0)
+\sum_0^\infty \frac{\lambda^n}{n!} \;
 \partial_\lambda^n \tG_{{\bf H}_\lambda}(t,t_0)\Big|_{\lambda=0}.
\end{equation}
Noticing that
$\tG_{{\bf H}_\lambda}(t_{i+1},t_{i})\big|_{\lambda=0}
=\tG_{\bf B}(t_{i+1},t_{i})$,
we obtain
\begin{equation}
   \tG_{{\bf H}_\lambda}(t,t_0)
 =  \tG_{\bf B}(t,t_0)
   +\sum_{n=1}^\infty
    \lambda^n\;\prod_{i=1}^n \Bigg(
    \int_{t_0}^{t} dt_i\;
    \tG_{\bf B}(t_{i+1},t_{i})\; {\bf A}(t_{i})
    \Bigg)
    \tG_{\bf B}(t_1,t_0),
\end{equation}
We may set $\lambda=1$ at this point.

Identifying ${\bf A}={\bf P}^A$,
$\tG_{\bf H}=\tG_{P^A+P^B}$ and
$\tG_{\bf B}=\tG_{P^B}$
we obtain more familiar identity%
\footnote{As previously we define 
  $\tG_X(t_i,t_j)={\bf G}_X(t_i,t_j)\theta_{t_i>t_j}$. }
\begin{equation}
   \tG_{P^A+P^B}(t,t_0)
 =  \Bigg\{ {\bf I}
   +\sum_{n=1}^\infty
    \prod_{i=1}^n \bigg(
    \int_{t_0}^{t} dt_i\;
    \tG_{B}(t_{i+1},t_{i})\; {\bf P}^A(t_{i})
    \bigg)\Bigg\}
    \tG_{B}(t_1,t_0),
\end{equation}
which leads immediately to eqs.~(\ref{eq:solumatr}) and (\ref{eq:Evolsolu}).
In this way we have completed the {\em second} proof of eq.~(\ref{eq:Evolsolu}) --
this time without any reference to backward evolution operator ${\bf G}^{-1}$.

\subsection{Straightforward derivation}
In addition to two elegant proofs of eq.~(\ref{eq:Evolsolu})
presented in the previous Sections,
we include in Appendix third proof,
which relies on a rather straightforward method
-- it starts from eq.~(\ref{eq:solumatr2}) and through
tedious reorganization of the sums over flavour indices
(change of the summation order) and relabeling of 
the variables transforms it into eq.~(\ref{eq:Evolsolu}).
The advantage of this third proof is that relation between
integration and summation variables in both formulas is exposed
in a manifest way.
This might be useful in the construction of the exclusive MC model
of the parton shower type.

\section{Discussion}
We are fully aware, of course,
that all three derivations of eq.~(\ref{eq:Evolsolu}),
shown in this work represent a well established
mathematical formalism, very similar to that in use in the Quantum Mechanics,
theory of Markovian processes and renormalization group in 
the Quantum Field Theory.
We did not add much to the development of the corresponding area of mathematics.
Rather, our main aim was to customize 
this known formalism to the specific needs of solving 
the QCD evolution (also numerically),
such that solution of eq.~(\ref{eq:Evolsolu})
and the other similar ones are obtained in an effortless and rigorous way.
Having all this in mind, let us
comment on certain selected aspects
of the presented formalism,
on their possible refinements, extensions and applications.
We shall concentrate mainly on two points:
\begin{itemize}
\item Extension to beyond-DGLAP evolutions in QCD, like CCFM and others.
\item Possible application in the Markovian MCs and the related question
      of the momentum sum rules and normalization of PDFs.
\end{itemize}

\subsection{Extensions beyond DGLAP evolution}
In our definitions of the evolution of PDFs
eqs.~(\ref{eq:Evolequ}-\ref{eq:Evolsolu}) and the rest of the paper
we have restricted ourself to DGLAP type~\cite{DGLAP} evolution,
leading-logarithmic (LL) version or its next-to-LL extensions.
This restriction is however inessential and the validity of our
derivations can be extended to a more general evolution equation
\begin{equation}
\label{eq:genevoleq}
  \partial_t D_k(t,x)
 = \sum_j \int_x^1 du\; \Keu_{kj}(t,x,u) D_j(t,u),
\end{equation}
in which the dependency of the generalized
kernel $\Keu(t,x,u)$ is more general than only through the ratio $z=x/u$.
The above more general evolution equation is used for instance
in the CCFM-type models%
\footnote{It is also closer to the spirit of of the
          parton shower MC and unintegrated PDFs.}
of PDF \cite{CCFM}.
The DGLAP case of eq.~(\ref{eq:Evolsolu}) is obviously covered
by eq.~(\ref{eq:genevoleq}), with the following identification
\begin{equation}
  \Keu_{kj}(t,x,u) = \frac{1}{x}{\Peu}_{kj}\left(t,\frac{x}{u}\right)
  = \frac{\alpha(t)}{\pi} \frac{1}{x} P_{kj}\left(t,\frac{x}{u}\right).
\end{equation}
The compact matrix notation used in the time-ordered exponentials
can easily accommodate multiplications of the kernels $\Keu(t,x,u)$
such that all relevant algebra in the previous sections remains unchanged.
Let us only indicate how the product of two kernels gets redefined
\begin{equation}
   \big( {\bf P}(t_2){\bf P}(t_1)\big)_{kj}(x,u)=
   \sum_{j'} \int_x^1 du'\; \Keu_{kj'}(t_2,x,u') \Keu_{j'j}(t_1,u',u).
\end{equation}
The reader can easily verify that the rest of the compact matrix algebra
in our derivations remains unchanged.

\subsection{Sum rules and Markovianization}
As already mentioned, results of this work were instrumental
in the modelling QCD evolution using {\em non-Markovian} type
Monte Carlo techniques in 
refs.~\cite{Jadach:2005bf} and \cite{Jadach:2005rd}.
The corresponding MC programs simulate
DGLAP and CCFM class evolutions.

However, the solution eq.~(\ref{eq:Evolsolu}) may be also used
to construct an interesting example of the Markovian MC in which
single step in the Markovian chain is a Markovian process of its own.
Without going into fine details, let us indicate how this can be done.
To this end we have to invoke momentum sum rule
\begin{equation}
  \sum_k \int_0^1 dx\; xD_k(t,x) =0
\end{equation}
and reorganize slightly eq.~(\ref{eq:Evolsolu}).
Staying for simplicity with the DGLAP case (LL and beyond)
the above sum rule determines virtual part of the kernel
\begin{equation}
  \Peu^{\delta}_{jj}(t,\veps)=
  \sum_k \int_0^1 dz\; z\Peu^\Theta_{kj}(t,z).
\end{equation}
The above is used to set up properly Markovian MC and in particular
to split Sudakov formfactor $\Phi_k(t)$ into bremsstrahlung part and the rest
(flavour changing part)
\begin{equation}
\begin{split}
& {\Phi_{k}(t,t_0)} ={\Phi^{A}_{k}(t,t_0)}+{\Phi^{B}_{k}(t,t_0)},
\\&
   {\Phi^{B}_{k}(t,t_0)} = 
  \int_{t_0}^t  dt'\; \int_0^1 dz\; z\Peu^\Theta_{kk}(t',z).
\\&
   {\Phi^{A}_{k}(t,t_0)} = 
  \int_{t_0}^t  dt'\; 
 \sum_{j,\; j\neq k} \int_0^1 dz\; z\Peu^\Theta_{jk}(t',z).
\end{split}
\end{equation}

By means of pulling out $\Phi^{A}_{k}(t)$ and multiplying both sides of
eq.~(\ref{eq:Evolsolu}) by $x$, we obtain the following
formula suitable for a Markovian MC
\begin{equation}
\label{eq:Evolsolu2}
\begin{split}
xD_k(t,x) =&
     \int\limits_0^1 dz'\; dx_0\; 
     \Ucal_{kk}^B(t, t_{0},z')\;
     x_0D_k(t_0,x_0)\delta(x-z' x_0)
  +\sum_{n=1}^\infty \;
      \int\limits_0^1 dx_0\;
\\&\times
   \sum_{{k_{n-1},\dots,k_{1},k_{0}}}
   \Bigg[ \prod_{j=1}^n \int\limits_{t_0}^t dt_j\;
     \Theta(t_j-t_{j-1}) \Bigg]\;
   \int\limits_0^1 dz'_{n+1}\;
   \Bigg[ \prod_{i=1}^{n} 
     \int\limits_0^1 dz'_i\;dz_i\;
     \Bigg]\;
\\&\times
   \Ucal_{kk}^B(t,t_n,z'_{n+1})\;
   \Bigg[ \prod_{i=1}^n
     e^{-\Phi^{A}_{k_{i-1}}(t_i,t_{i-1})}
     z_i\Peu_{k_i,k_{i-1}}^{A} (t_i,z_i)\;
     \Ucal_{k_{i-1}k_{i-1}}^B(t_i,t_{i-1},z'_i) \Bigg]\;
\\&\times
      x_0 D_{k_0}(t_0,x_0) 
      \delta\bigg(x- x_0 \prod_{i=1}^n z_i \prod_{i=1}^{n+1} z'_i
     \bigg),
   \;\;\; k_n=k,
\end{split}
\end{equation}
where
\begin{equation}
\Ucal_{kk}^B(t_1,t_0,z)
\equiv e^{\Phi^{A}_{k}(t_1,t_0)}
       z G_{kk}^B(t_1,t_0,z)
\end{equation}
obeys evolution equation of eq.~(\ref{eq:GAbrems}),
with the substitution 
$\Phi_{k}(t_i,t_{i-1})\to  \Phi^{B}_{k}(t_i,t_{i-1})$
and with both side multiplied by $z$.
The evolution operator $\Ucal$ obeys nice "unitarity" rule
\begin{equation}
  \int dz\; \Ucal_{kk}^B(t,t_0,z) =1
\end{equation}
for any $t$, $t\geq t_0$, in addition to the usual boundary condition
$\Ucal_{kk}^A(t_0,t_0,z)=\delta(1-z)$.

Eq.~(\ref{eq:Evolsolu2}) can be now used to define 
hierarchic (nested) Markovian Monte Carlo algorithm.
The normalized probability distribution of the forward
Markovian step in the flavour-changing 
upper level Markovian process reads:
\begin{equation}
\begin{split}
&\omega(t_i,x_i,k_i|t_{i-1},x_{i-1},k_{i-1})
 =\omega(t_i,z_i z'_i x_{i-1},k_i|t_{i-1},x_{i-1},k_{i-1})
\\&~~~~~~~~~~~~~~~~~~
 =(1-\delta_{k_ik_{i-1}})
      e^{-\Phi^{A}_{k_{i-1}}(t_i,t_{i-1})}\;
      z_i \Peu_{k_ik_{i-1}}^A (z_i)\;
      z'_i \Ucal^B_{k_{i-1}k_{i-1}}(t_i,z'_i | t_{i-1}),
\\&
\int_{t_{i-1}}^{\infty} dt_i \; \sum_{k_i} 
\int_0^1 d z_i \int_0^1 dz'_i\;\;
   \omega(t_i, z_i z'_i x_{i-1},k_i|t_{i-1},x_{i-1},k_{i-1})
\equiv 1.
\end{split}
\end{equation}
For the lower level bremsstrahlung process one may use standard
Markovian MC technique of ref.~\cite{Golec-Biernat:2006xw}.

Let us discuss selected details of the above scenario.
Here, all $k_i$ and $t_i$, $i=1,2,...n$ can be generated before
generation of any $z$-variables in a separate Markovian algorithm
with the stopping rule being the usual condition $t_{n+1} \geq t$.
(See ref.~\cite{Golec-Biernat:2006xw} for more details on the
the Markovian class MC algorithms.)
Variables $z_i$ of the flavour-changing kernels can also be generated
at this early stage.

The interesting question is: how and when do we generate
$z'_i$ according to gluonstrahlung operator
$\Ucal^B_{k_{i-1}k_{i-1}}(t_i,z'_i | t_{i-1})$?
If we have known an analytical (even approximate)
representation of this function,
or its precise value from the look-up tables,
then we could readily generate them before entering into
MC simulation of the bremsstrahlung subprocess.
That would lead us to the use of the constrained MC of 
refs.\cite{Jadach:2005bf,Jadach:2005rd,zinnowitz04} for the bremsstrahlung segments.
Alternatively, $z'_i$ may come out from a separate
Markovian MC module simulating gluonstrahlung sub-process
starting at $t_{i-1}$ and stopping at $t_i$,
with the normalized probability distribution of single Markovian
step defined in ref.~\cite{Golec-Biernat:2006xw}.
In this latter case we would have simulated in the MC
a hierarchic system of Markovian processes,
with the master flavour-changing Markovian process and many Markovian subprocesses,
each of them implementing pure brems\-strahlung, flavour conserving, emissions.

The above hierarchic Markovian MC scheme, although quite interesting,
seems to have no immediate practical importance.
However, it may find applications in some future works.


\section{Summary}
The basic aim of this paper is to provide solid technical foundation
to other works in the area of the Monte Carlo simulation of the evolution
of PDFs and parton shower in QCD.
Our basic result is the solution of the evolution equation
of eq.~(\ref{eq:Evolsolu}), which was proved algebraically using three method.
Its primary application is construction of the constrained 
MC algorithms of ref.~\cite{Jadach:2005bf}.
In addition, we also describe hypothetical application of such a solution
in the Markovian MC algorithm.
Although we are aware of many interesting relation of the discussed problems
and solutions to other areas in physics, 
we did not attempt to elaborate on that too much, in order to keep
the paper compact and transparent.
Let us mention also, that our solution can be used many times
leading to a nested structure with several levels of the hierarchy.

\vspace{10mm}
\noindent
{\bf\large Acknowledgments}

\noindent
We would like to thank W. P\l aczek  for  useful discussions 
and reading the manuscript.
We thank for warm hospitality of the CERN PH/TH division were part
of this work was done.

\newpage

\appendix
{\bf\LARGE Appendix}\\
{\section{Combinatorial proof}
We are going to show how to transform non-hierarchic
solution in eq.~(\ref{eq:SolIter}) (with resummed virtual corrections)
into hierarchic solution in eq.~(\ref{eq:Evolsolu})
(with resummed gluonstrahlung) using straightforward method
of changing summation order and relabeling integration variables.

The critical point in isolating two levels in the evolution,
flavour-changing transitions and gluonstrahlung,
will be the change of the summation order
in eq~(\ref{eq:Evolsolu}), such that one is able to resum separately
the pure brems\-strahlung segments obeying $k_i=k_{i-1}$.
These segments will form (many) functions $G^A_{kk}$, as defined
in eq.~(\ref{eq:GAbrems}).
The corresponding transformation of the summation order (indexing)
looks schematically as follows
\begin{equation}
\begin{split}
&\sum_{n=0}^\infty 
    \sum_{k_{n-1}\dots,k_{1}k_{0}} t_{k_nk_{n-1}\dots k_{1}k_{0}}=
\sum_{n=0}^\infty
\sum_{{k_{n-1}\dots,k_{1}k_{0}}\atop %
      {k_{n}\neq k_{n-1}\neq\dots \neq k_{1}\neq k_{0}}}
\\& 
\sum_{j_n,j_{n-1}\dots j_0=1}^\infty
     t_{ k_{n}^{(j_n)}\dots k_{n}^{(2)}k_{n}^{(1)}%
         k_{n-1}^{(j_{n-1})}\dots k_{n-1}^{(2)}k_{n-1}^{(1)}%
         \dots\dots%
         k_{1}^{(j_1)}\dots k_{1}^{(2)}k_{1}^{(1)}%
         k_{0}^{(j_0)}\dots k_{0}^{(2)}k_{0}^{(1)}}
\end{split}
\end{equation}
where we have $k_{r}^{(j_r)}=\dots =k_{r}^{(2)}=k_{r}^{(1)}$
and the purpose of the upper index in this context is simply to show 
that the same index $k$ is repeated $j_r$ times.
On the other hand variables
$z^{(m)}_r$ and $\tau^{(m)}_r,\quad r=1,2,\dots,n,\quad m=1,2,\dots,j_r$
are truly different (independent), with the upper index
truly differentiating them.

The aim is now to show that one can
factorize out the functions $G^A_{kk}$
and identify precisely the remaining functions and integrations.
Employing the above index transformation
in the product of the $\Peu$-functions we obtain
\begin{equation}
\begin{split}
\{ \Peu_{k_nk_n}(t_{j_n}^{(n)},z_{j_n}^{(n)})\dots 
       \Peu_{k_nk_n}(t_{2}^{(n)},z_{2}^{(n)})\} 
         &\Peu_{k_nk_{n-1}}(t_{1}^{(n)},z_{1}^{(n)})
\\ \dots &
\\ \dots\;\dots \Peu_{k_1k_1}(t_{2}^{(2)},z_{2}^{(2)})\}
         &\Peu_{k_2k_1}(t_{1}^{(2)},z_{1}^{(2)})\; 
\\ \{\Peu_{k_1k_1}(t_{j_1}^{(1)},z_{j_1}^{(1)})\dots 
     \Peu_{k_1k_1}(t_{2}^{(1)},z_{2}^{(1)}) \}&\Peu_{k_1k_0}(z_{1}^{(1)},z_{1}^{(1)}) 
\\ \{ \Peu_{k_0k_0}(t_{j_0}^{(0)},z_{j_0}^{(0)})\dots 
            \Peu_{k_0k_0}(t_{2}^{(0)},z_{2}^{(0)})\;\;
         &\Peu_{k_0k_0}(t_{1}^{(0)},z_{1}^{(0)}) \},
\end{split}
\end{equation}
where curly bracket embrace the diagonal elements $\Peu_{kk}$,
to be collected into the $G^A_{kk}$-functions; the remaining,
nondiagonal ones, are now clearly isolated.

Each $\Peu_{k_i,k_{i-1}}$, $k_i\neq k_{i-1}$ in eq.~(\ref{eq:SolIter})
is accompanied by an exponential factors.
All of them
(including the first one which does not belong to any $\Peu$)
are now reorganized as follows:
\begin{equation}
\begin{split}
&    e^{\Phi_{k_n}(t,t^{(n)}_{n})}
\\&  e^{\Phi_{k_n}(t^{(n)}_{n},t^{(n)}_{j_n,1})}
   \dots
     e^{\Phi_{k_n}(t^{(n)}_{2}  ,t^{(n)}_{1})}
     e^{\Phi_{k_{n,1}}(t^{(n)}_{1}  ,t^{(n,1)}_{j_{n,1}})}
\\&   \dots
\\&    e^{\Phi_{k_1}(t^{(1)}_{j_1},t^{(1)}_{j_1,1})}
   \dots
     e^{\Phi_{k_1}(t^{(1)}_{2}  ,t^{(1)}_{1})}
     e^{\Phi_{k_0}(t^{(1)}_{1}  ,t^{(0)}_{j_1})}
\\&  e^{\Phi_{k_0}(t^{(0)}_{j_1},t^{(0)}_{j_1,1})}
   \dots
     e^{\Phi_{k_0}(t^{(0)}_{2}  ,t^{(0)}_{1})}
     e^{\Phi_{k_0}(t^{(0)}_{1}  ,t_0)}
\\=
&  e^{\Phi_{k_n}(t, t^{(n)}_{1})}\; 
   e^{\Phi_{k_{n,1}}(t^{(n)}_{1} ,t^{(n,1)}_{1})}\;
   e^{\Phi_{k_1}    (t^{(2)}_{1} ,t^{(1)}_{1})}\;
   e^{\Phi_{k_0}    (t^{(1)}_{1} ,t_0)},
\end{split}
\end{equation}
where all factors entering into products in $G^A_{kk}$-functions
are shown inside the curly brackets. 
Together with the flavour-changing $\Peu$'s,
the above exponential form-factors look as follows:
\begin{equation}
\begin{split}
&  e^{\Phi_{k_n}(t, t^{(n)}_{1})}\quad
  \Peu_{k_nk_{n,1}}(z_{1}^{(n)}) e^{\Phi_{k_{n,1}}(t^{(n)}_{1} ,t^{(n,1)}_{1})}
\\& ~~~~~ \dots ~~~~~
   \Peu_{k_2k_1}(z_{1}^{(2)})    e^{\Phi_{k_1}    (t^{(2)}_{1} ,t^{(1)}_{1})}~~~~
   \Peu_{k_1k_0}(z_{1}^{(1)})    e^{\Phi_{k_0}    (t^{(1)}_{1} ,t_0)}.
\end{split}
\end{equation}
The other diagonal $\Peu$'s will enter into $G^A_{kk}$-functions.

In this rather sketchy way we have shown
that indeed  eq.~(\ref{eq:SolIter}) can be transformed
into eq.~(\ref{eq:Evolsolu}) by means of the straightforward reorganization
of multiple sums and integrals.
(More systematic proof would require completing
mathematical induction with respect to $n$.)

\bibliographystyle{utphys_spires}
\providecommand{\href}[2]{#2}

\end{document}